# Recording Reusable and Guided Analytics From Interaction Histories

Nam Wook Kim*

LG Electronics

**ABSTRACT**

The use of visual analytics tools has gained popularity in various domains, helping users discover meaningful information from complex and large data sets. Users often face difficulty in disseminating the knowledge discovered without clear recall of their exploration paths and analysis processes. We introduce a visual analysis tool that allows analysts to record reusable and guided analytics from their interaction logs. To capture the analysis process, we use a decision tree whose node embeds visualizations and guide to define a visual analysis task. The tool enables analysts to formalize analysis strategies, build best practices, and guide novices through systematic workflows.

**Index Terms**: H.5.2 [Information Systems]: Information Interfaces and Presentation—User Interfaces

## 1 INTRODUCTION

Organizations gather massive amounts of data and in turn construct large data warehouses to understand key aspects of their business operations. To support analysis of large and complex data sets, the use of visual analytics tools is becoming increasingly important. Visual analytics involves more than merely generating and manipulating visualizations; it also needs to support the analysis process [2]. In order to facilitate communication of what users have found and how they have found it, the paths of investigation and reasoning behind the discovery have to be preserved and shared [3].

There has been a growing interest in understanding and recording a user's analysis process in visual data analysis. Previous research has shown that interaction logs can be used to recover the analysis process and can serve as an effective memory aid for recalling analysis strategies [1, 3]. Other studies have also looked at the value of supporting users to externalize their analysis process into artifacts by using diagrams [5, 6]. Perer and Shneiderman recently presented *systematic yet flexible* design goals to support the analysis process by decomposing it into systematic steps for guided exploration [4]. Although the previous systems and approaches are proven to be innovative and effective, their emphasis differs from our goal in that none of these systems fully addressed our question of how the analysis process can be recovered from interaction logs and in turn be reproduced to guide novices.

This work-in-progress proposes a visual analysis tool that enables analysts to collaboratively record analysis strategies from their interaction logs using a decision tree. The end product is an analysis workflow template that can be replayed on new data to guide novices through the decision-making steps of the tree. We have implemented a prototype in a manufacturing domain in the hope that the analysis method employed by expert analysts, to find causal factors for manufacturing defects, can be shared with less educated operators in small production sites. This work contributes to the research goal of building a tool that enhances the dissemination of knowledge discovered through complex visual analysis.

## 2 DESIGN DECISIONS

Thomas and Cook recommend that researchers in visual analytics develop tools to support collaborative analytic reasoning among people with different levels of expertise and backgrounds and to communicate the discovered knowledge through the use of an appropriate visual metaphor [7]. Our design decisions attempt to comply with this guideline by using a tree representation to capture an analysis process that is built through collaboration and shared among experts and novices. We applied the design decisions to build a system that comprises three components: an analysis view, an analytics editor and guided analytics (Fig.1).

### 2.1 Analysis View

The analysis view consists of a suite of interactive visualizations which support changing visual mapping parameters such as changing axes in a bar chart (Fig. 1a). We currently support four conventional charts (bar chart, line chart, scatter plot, and histogram) and two domain-specific charts (PCB defect and nozzle center data visualizations). Statistics such as regression line, moving average, weighted moving average and the Pareto Principle are also provided to aid in-depth analysis of production data. Measures and dimensions of the underlying data cubes are used to specify visual mapping parameters. For example, setting x-axis to 'lines', y-axis to 'defect rate', and stack to 'chip mounters' in the bar chart view will draw defect rates of the lines stacked with chip mounters. An analyst interacts with the analysis view to construct hypotheses, experiment with analysis methods and discover knowledge. In addition, the interaction history of changing visualization states is saved for later recall of the analysis process and can be shared among analysts for collaboration (Fig. 1b).

### 2.2 Analytics Editor

The analytics editor is an authoring tool for creating a decision tree to encode an analysis strategy. Each node in the tree defines an analysis task by integrating visualizations from the analysis view along with the guide to describe the task and provide instructions on how to analyze the data. In addition, decisions being made in the task are modeled using a group of radio buttons. Interaction logs from the analysis view are used to recover the analysis strategy behind the discovered knowledge and can be referred to in order to record a reusable analysis process.

A diagramming tool is provided for editing the decision tree (Fig. 1c). We categorize a node into seven types of tasks, involving analysis of chip mounters, PCBs, electronic materials, mounting location, devices (head, nozzle, feeder) that are key aspects of production data, and control setting. The control setting is adjusted to control specification of the causal factors (e.g., material pickup location of a nozzle) to prevent future defects, forming a closed-loop system. In addition, a group of analysts can work together to build the decision tree by branching out of existing nodes.

\* e-mail: namwkim85@gmail.com

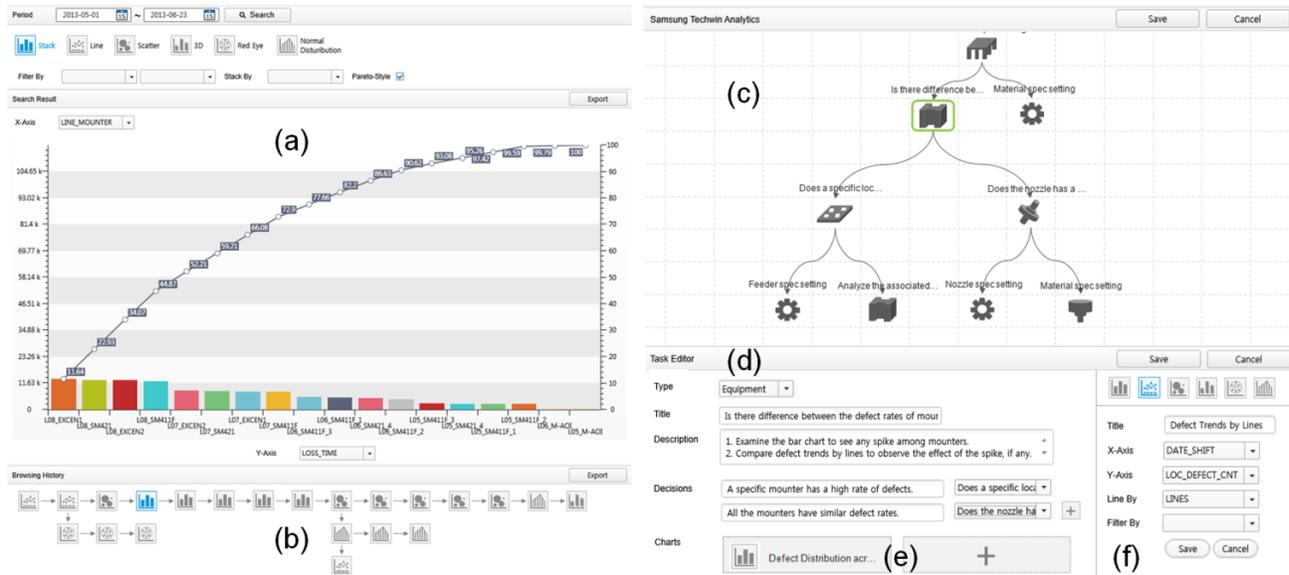

Figure 1: An initial prototype. (a) shows the current visualization state. (b) shows the history of changing visualization states. (c) shows the diagramming interface for editing the decision tree. (d) shows the task editor for editing an analysis task. (e) shows an area for adding visualizations while (f) shows the editing section to manually specify a visualization.

For each node, the task editor is used to define an analysis task (Fig. 1d). A task has guides, decisions and visualizations. To reduce the burden in manually specifying a visualization to be used in the task, simply dragging a visualization state from the interaction history to the task editor (from Fig 1b to Fig 1e) transfers visualization parameters (the visualized data is not transferred). A user also can change, or manually create, the specification as shown in Fig. 1f (e.g., specifying a dynamic filter based on the selected object in the previous task).

## 2.3 Guided Analytics

A resulting analytics defines an analysis workflow template which becomes live with real data. The embedded visualizations are parameterized by measures and dimensions of the underlying data cubes and their actual graphics are generated with the real data. A user can see the overview of the analysis process (tree), step through the tasks (nodes), and see the progress through the high-lighted paths of completed tasks (sub-tree). The process is shared with others and remaining tasks can be delegated to them, if necessary. An example of the task instantiated with the live production data is shown in Fig. 2.

## 3 CONCLUSION

In this poster, we have demonstrated a novel system that enables analysts to create a reusable analysis process from interaction logs. However, the cognitive burden involved in manually creating the process is not negligible. The future work should address this issue by improving the automatic leveraging of interaction logs. To this end, interaction logs should reflect the representation of the reasoning process in an analyst's mind, not just produce snapshots of events and states.

Our work assumes that the database schema does not change in order for the analysis strategy to be transferrable. Future studies could pursue different designs that are able to incorporate disparate data sources. In addition, while we used a tree representation to focus on decision-making problems, the creation of other representations, such as a network, would be an interesting avenue for future work.

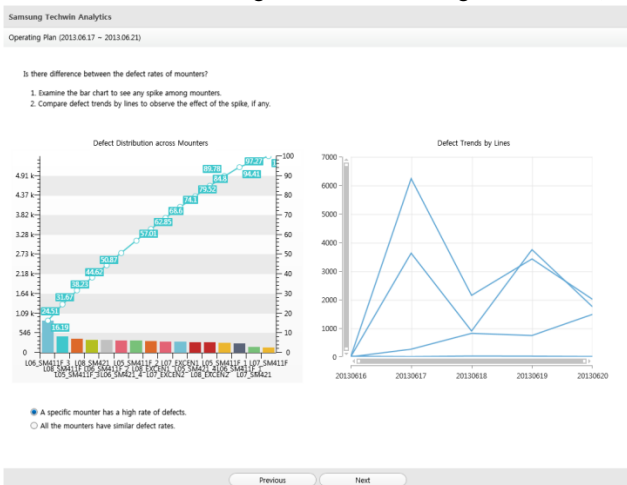

Figure 2: A task page example. Visualization parameters are used to fetch the data to generate graphics.